\documentstyle[preprint,epsfig,aps]{revtex}

\begin{document}

\centerline{\bf Signals of phase transition in atmospheric showers} 

\centerline{ C. Pajares, D. Sousa, and R.A.~V\'azquez}
 
\centerline{\it Departamento de F\'\i sica de Part\'\i culas, Universidade de
Santiago}
\centerline{\it E-15706 Santiago de Compostela, Spain}

\begin{abstract}
We discuss collective effects like percolation, quark gluon plasma, or string 
fusion and their effect on the longitudinal development of high energy showers 
in air. It is shown that iron--air showers could develop more slowly than 
expected and produce the observed change in the slope of $X_{\mbox{max}}$ at 
the highest energies. 
\end{abstract}

It is expected that the Relativistic Heavy Ion Collider (RHIC) and the 
Large Hadron Collider (LHC) will reach regions where the energy density will be
higher than the critical value of a few GeV/fm$^3$ predicted by lattice QCD 
calculations. This critical energy density corresponds to the new phase(s) of 
deconfinement of the quark gluon plasma or (and) of restoration of chiral 
symmetry. The energies of RHIC ($\sqrt s = 200$ AGeV) and LHC ($\sqrt = 7000 $
AGeV) correspond, for Ag--Ag collisions, to laboratory energies slightly 
higher than 10$^{15}$ and 10$^{18}$ eV, which are in the range of typical 
very high energy cosmic rays. Unfortunately in cosmic rays projectile and 
target are not heavy nuclei. The target is always Air and the heaviest 
projectile is Iron. Nevertheless, it would be worth to investigate whether 
cosmic ray collisions in the atmosphere can produce an hadronic phase 
transition and which are the main consequences related to the development of 
cosmic ray cascades.

Recently, it has been pointed out that the confined to deconfined phase 
transition could take place as a percolation of the strings formed in a 
nucleus-nucleus collision \cite{Armesto,Satz}. For a given collision and a 
given degree of the centrality of the collision there is a defined available 
total area in the impact parameter space. In this space, the strings exchanged 
between the partons of projectile and target are seen as circles of radius 
$r$ inside the area. As the energy and the size of the projectile and target 
increase, more and more strings overlap. The cluster of several overlapping 
strings form a new string with an energy momentum given by the sum of the 
energy momentum of the old strings. Also, the new string will have a higher 
tension. Above a critical density of strings, percolation occurs, so that paths
of overlapping circles are formed through the whole collision area. The 
percolation is a second order phase transition on the nuclear scale. The 
percolation threshold $\eta_C$ is related to the critical density of circles 
$n_C$ by the expression
\begin{equation}
\eta_C = \pi r^2 n_C
\end{equation}
$\eta_C$ has been computed using different methods. All the results are in the
range $\eta_C = $ 1.12 -- 1.18 \cite{eta1,eta2,eta3,eta4}. Taking the 
reasonable value of $r \sim $ 0.2  fm, it is obtained a critical density of 
strings, $n_C = $ 8.9 -- 9.3 strings/fm$^3$. All the AGS and SPS experiments 
will give rise to string density values below the critical density. Only 
central Pb-Pb collisions at SPS energies are slightly above 9.5 
strings/fm$^3$. At RHIC energies it is expected to reach the critical density 
for Ag-Ag and at LHC for S-S collisions.
 
For central Fe-Air collisions we have used a MonteCarlo code based on the 
quark gluon string model (QGS) to compute the number and the density of 
strings at $E_{lab} = 10^{17} $ eV obtaining a value very close to the 
critical one. Therefore in the range $10^{17} - 10^{20} $ eV, we will expect 
the phase transition to occur. In table \ref{table1} we can see the number and 
density of strings as calculated by our MonteCarlo for different energies, 
projectiles and targets for central collisions. Although, iron cosmic ray 
collisions are not central but minimum bias, as $\pi R^2_{\mbox{Air}}/( 
\pi (R_{\mbox{Air}} + R_{\mbox{Fe}} )^2 ) \sim 1/6.6 $, the number of central 
collisions is not negligible.

The main effects of phase transition related to cosmic ray cascades are: a) 
dumping of the multiplicity, b) enhancement of the cummulative effect c) 
enhancement of heavy flavour production d) increase of the inelasticity.

The dumping of the multiplicity is a consequence of the overlapping of the 
strings forming a new one, in such a way, that the effective number of 
independent strings (clusters) is reduced compared to the number of original 
strings. The fragmentation of the new string can give rise to particles 
outside the kinematical limits of a nucleon--nucleon collision (the so called 
cummulative effect) and the shape of the x-distribution would be harder than 
the case of fragmentation of the original strings. 

The new strings have also a larger tension due to the higher colour of the 
grouped partons located at the end of the string which is the SU(3) non 
abelian sum of the colour of the partons of each original string. This larger 
tension produces more strange, baryon and heavy flavour particles. The main 
effect on the cascade of this is a further reduction of the multiplicity. The 
conservation of energy implies that an enhancement of heavy flavour means a 
suppression of pions. 

\begin{figure}
\epsfxsize=10cm
\begin{center}
\mbox{\epsfig{file=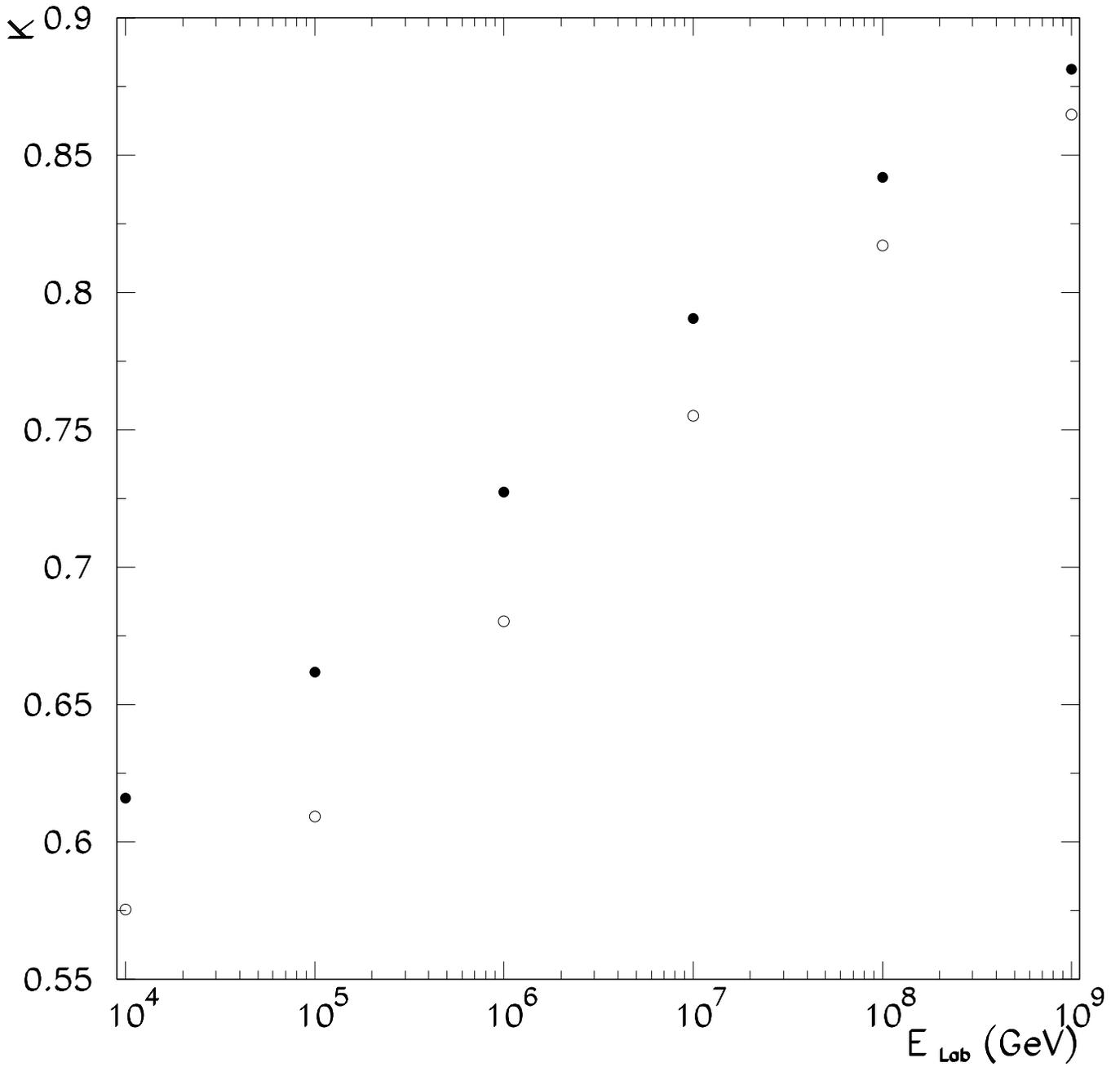}}
\end{center}
\caption{Inelasticity of Fe--Air collisions as a function of the total lab 
energy for the model with (filled circles) and without fusion (open circles).} 
\label{inelasticity}
\end{figure}

Finally, the overlapping of strings produces an increasing of the stopping 
power, {\it i.e.} the inelasticity. To see further in this point, we have 
computed in our MonteCarlo code the inelasticity for central Fe-Air in the 
QGS model at different energies and compared the results to the quark string 
fusion model (QGSF). The difference between both models is the possible fusion
of two strings into one if the original ones are close to 0.4 fm. In fig. 
\ref{inelasticity} it is plotted the inelasticity as a function of the energy 
for both cases (with and without fusion), showing the mentioned rise of the 
inelasticity. In the case of percolation of strings the rise would be larger.

Contrary to what could be thought, this rise of the inelasticity does not 
imply a shortened of the shower depth of maximum, $X_{\mbox{max}}$. In fact, 
the rise of the inelasticity is compensated with the reduction of the 
multiplicities and with a harder $x$ distribution in such a way that 
$X_{\mbox{max}}$ rises significatively. Indeed, an identification of the 
primary as iron together with a behaviour of the cosmic ray cascade more 
similar to a typical proton cascade could be a good signal of quark gluon 
plasma.

In order to compute the effects of string fusion on the atmospheric cascade 
development, we have built a toy Monte Carlo code to simulate hadronic 
showers in air. We have used as the hadronic generator for our code a 
modified version of the Hillas algorithm \cite{Hillas} as given by Lipari 
\cite{Lipari}. This algorithm allows to choose the hadronic cross sections 
$d n/dx$ by varying a set of parameters. The algorithm has the advantage of 
being flexible, fast and easy to program. The electromagnetic cascade was 
simulated using the A approximation for the electromagnetic cross sections, 
{\it i.e.} we will consider only bremsstrahlung and pair production. Since the
hadronic component of the shower is dominated by the production of pions we 
will consider only the following hadronic reactions: $p Air \rightarrow p$, 
$p Air \rightarrow \pi$, $\pi Air \rightarrow \pi$.

A complete Monte Carlo computing all the resonances and the hadronic cross 
sections is out of the scope of this work and is currently under research 
\cite{prep}. 

The hadronic cross sections simulated by this program and the set of parameters
which define an hadronic model are shown in ref. \cite{Lipari}. This model
has a set of 5 free parameters to change the shape of the hadronic cross 
sections. In ref. \cite{Lipari} the set of parameters are called $K_p,P^*,P_a
,P_b$, and $P_d$. For proton air collisions, $K_p$ defines the inelasticity of 
the leading proton and $P^*$ changes the charged pion rapidity distribution. 
$P_d$ defines the fraction of diffractive pions produced in pion air 
collisions and $P_a, P_b$ both change the shape and total multiplicity of pion
production in pion air collisions. The original algorithm due to Hillas, which
was implemented in the program MOCCA, had these parameters taken to be 
constants given by $K_p=1/2, P^*=0, P_d=1/2, P_a=1/2, P_b=0$.
We have checked that our code reproduces the results of the more realistic 
MonteCarlo, Aires \cite{Aires}, which uses the original Hillas algorithm.

By changing the value of the above parameters we can effectively simulate 
fusion. We can mimic the main effects that are expected from the fusion of 
strings by reducing the multiplicity and increasing the inelasticity of the 
proton and of the pion. 

Both modifications compete in opposite directions. A reduction on the 
multiplicity of pions will produce less pions in the first stages of the 
shower, but due to energy conservation they will be more energetic. As a 
consequence the shower will develop more slowly and the maximum will be 
reached at a greater depth. Increasing the inelasticity has the opposite 
effect. If the proton looses more energy in the first interactions, the shower
will develop faster and the maximum will be shortened. The net effect will be 
a combination of the two, but the change in multiplicity dominates the shower 
maximum. With very simplistic assumptions we have estimated that the maximum 
change is given by:
\begin{equation}
\frac{\Delta X_{\mbox{max}}}{ X_{\mbox{max}}} \sim - \frac{\Delta N}{2 N} - 
\frac{\Delta K}{10 K} 
\end{equation}
where $\Delta X_{\mbox{max}}$, $\Delta N$, and $\Delta K$ are respectively the 
modification of the shower maximum, multiplicity, and inelasticity. This means 
for instance that decreasing the multiplicity 100 \% will increase the depth 
of maximum a 50\%. Instead, increasing the inelasticity 100~\% will reduce the 
maximum of depth only a 10\%. This numbers are only order of magnitude 
estimations and depend on the specific details of the hadronic models chosen.
We can safely expect that the decreasing of the maximum depth of the shower
due to the increasing of the inelasticity is negligible compared to the 
increasing of $X_{\mbox{max}}$ due to the decreasing of the multiplicity. 

\begin{figure}
\epsfxsize=10cm
\begin{center}
\mbox{\epsfig{file=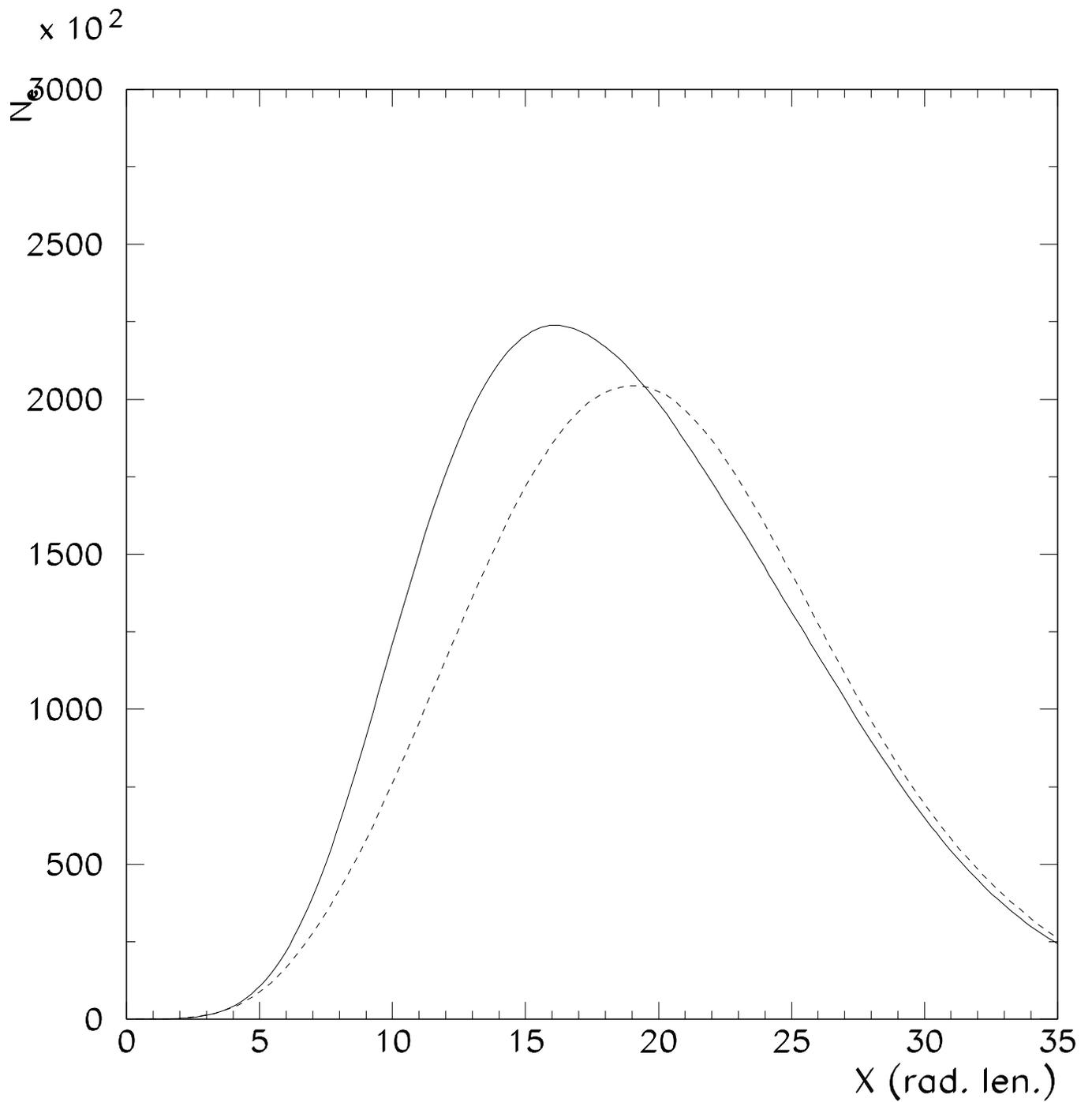}}
\end{center}
\caption{Number of electrons as a function of depth for showers of energy 
10$^7$ GeV for the two models: A (dashed line) and B (continuous line) as 
discussed in the text}
\label{shower}
\end{figure}

\begin{figure}
\epsfxsize=10cm
\begin{center}
\mbox{\epsfig{file=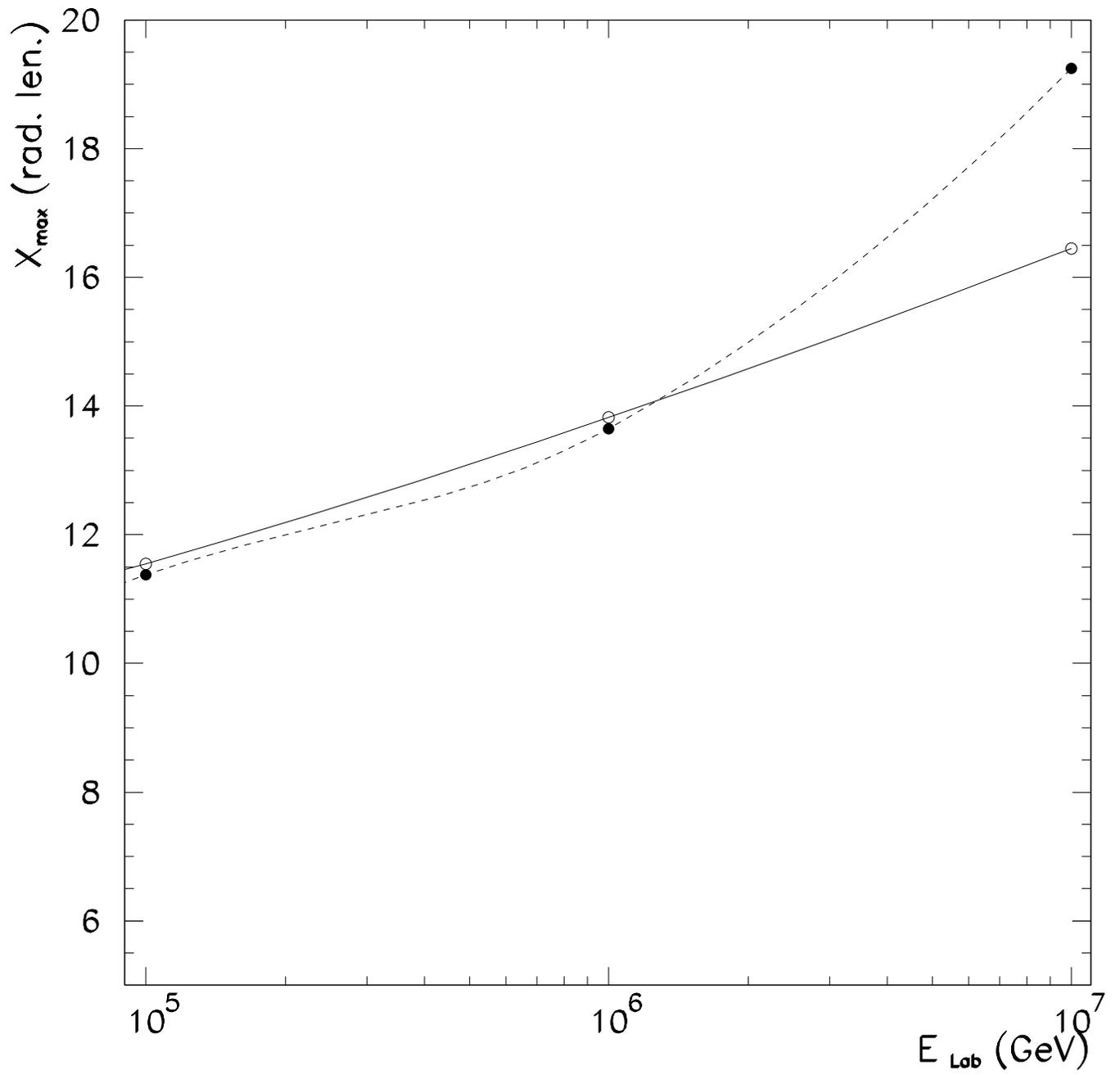}}
\end{center}
\caption{Depth of maximum as a function of the energy for the two hadronic 
models discussed in the text: A (dashed line) and B (continuous line).}
\label{xmax}
\end{figure}

We have used, to compute the effects of fusion, two hadronic models given by 
two sets of the above parameters. For the first model, model A, the set of 
parameters is chosen to mimic the fusion at the maximum degree as allowed by 
our set of parameters. {\it i.e.}, the multiplicity is reduced the maximum 
possible and the inelasticity is increased at the maximum. For the model B, the
opposite effect is selected. We set the inelasticity at the minimum and the 
multiplicity at the maximum possible with our set of parameters. In fig. 
\ref{shower}, we can see the shower profile of the two models for an initial 
energy of $E_0= 10^6$ GeV. We can see clearly the different shape and shower 
maximum for the two models. As expected, model A has a deeper maximum (and 
consequently a lower $N(X_{\mbox{max}})$). We have run our simulation for 
several energies from $10^4$ to $10^7$ GeV. And the results are summarized in 
fig. \ref{xmax}, where we plot the maximum of the shower as a function of the 
primary energy. We can see that at 10$^7$ GeV the shower maximum for the model
A is a 17 \% higher than the shower maximum for model B, and that this 
difference grows linearly with $\log E$. At low energy, as expected, the 
effect of fusion is not important. But we expect a large difference in the 
maximum of depth for energies of order 10$^{10}$ GeV. Extrapolating the given 
results we get 30 \% difference in the maximum of depth of the two models.

In conclusion we have shown that the fusion of strings predicts atmospheric 
showers with deeper shower maximum and that this effect is bigger enough to 
produce the observed change in the slope of $X_{\mbox{max}}$ versus energy 
observed by the cosmic ray experiments. This change of slope is usually 
interpreted as a change in the composition from heavy (iron) to light (proton)
and possibly indicating also a change in the origin of the cosmic rays. Here
we see that such a change could be given by a change on the hadronic 
interactions. If true, this may imply that the composition of cosmic rays does 
not change at the ankle and this would have profound implications for the 
origin of the most energetic cosmic rays. For Fe--Air central collision it is 
shown that the string density reached at $E_{\mbox{lab}} \sim 10^9$ GeV is over
the critical one and therefore the percolation of strings takes place. The 
cosmic ray physics overlaps with the physics of ultrarelativistic heavy ion 
accelerators like RHIC and LHC.

\acknowledgements

We thank C. Merino and E.G. Ferreiro for useful discussions. This work has 
been done under contract AEN96-1673 from CICYT of Spain. D.S. thanks Xunta de 
Galicia for financial support.

\begin{table}
\begin{displaymath}
\begin{array}{||c|c|c|c|c|c||}
\hline
\hline
\sqrt s \; \mbox{(AGeV)} &  & \mbox{p--p} & \mbox{S--S} & \mbox{Fe--Air} 
& \mbox{Pb--Pb} \\
\hline
19.4 & \mbox{number}  & 4.2  & 123  & 89    & 1145 \\
     & \mbox{density} & 1.3  & 3.5  & 4.42  & 9.5  \\
200  &                & 7.2  & 215  & 144   & 1703 \\
     &                & 1.6  & 6.1  & 7.16  & 14.4 \\
5500 &                & 13.1 & 380  & 255   & 3071 \\
     &                & 2.0  & 10.9 & 12.67 & 25.6 \\
\hline
\hline
\end{array}
\end{displaymath}
\caption{Number and density of strings as a function of energy for different
nucleus--nucleus collisions. The density is in units of strings/fm$^2$. At
$  s^{1/2} =$ 19.4, 200, and 5500 AGeV for Fe--Air, $E_{\mbox{lab}}$ is 1.1 
$10^4$, 1.1 $10^6$, and 8.6 $10^8$ GeV respectively.}
\label{table1}
\end{table}

\end{document}